\documentclass[9pt,twocolumn,twoside]{osajnl}
\journal{ol} % Choose journal (ao, aop, josaa, josab, ol, pr)

\setboolean{shortarticle}{true}
\title{Manipulating the transmission matrix of scattering media for nonlinear imaging beyond the memory effect}

\author[1]{Matthias Hofer}
\author[1,*]{Sophie Brasselet}

\affil[1]{Aix Marseille Univ, CNRS, Centrale Marseille, Institut Fresnel, F-13013 Marseille, France}

\affil[*]{Corresponding author: sophie.brasselet@fresnel.fr}

\begin{abstract}
The measurement of the Transmission Matrix (TM) of a scattering medium is of great interest for imaging. It can be acquired directly by interferometry using an internal reference wavefront. Unfortunately, internal reference fields are scattered by the medium which results in a speckle that makes the TM measurement heterogeneous across the output field of view. We demonstrate how to correct for this effect using the intrinsic properties of the TM. For thin scattering media, we exploit the memory effect of the medium and the reference speckle to create a corrected TM. For highly scattering media where the memory effect is negligible, we use complementary reference speckles to compose a new TM, not compromised by the speckled reference anymore. Using this correction, we demonstrate large field of view second harmonic generation imaging through thick biological media.
\end{abstract}

\setboolean{displaycopyright}{true}

\begin{document}

\maketitle

\section{Introduction}
\label{sec:intro}
As light propagates through a biological tissue its wavefront takes a random appearance and resembles a speckle. By shaping the incident wavefront at the entrance of a stable scattering medium, one can correct for this effect thanks to the deterministic nature of light propagation. An initial randomly scattered wave can be transformed into a focus by means of measuring the transmission matrix (TM) of the medium \cite{Popoff2011} or of optimization schemes \cite{Vellekoop2007}. The TM offers coherent control over a large field of view (FOV) through the medium, whereas optimization schemes are usually designed to focus light only in one specific point. Therefore, the TM serves as a valuable tool to study propagation characteristics of light through scattering media, such as polarization \cite{Tripathi2012,DeAguiar2017}, pulse broadening \cite{Andreoli2015,Mounaix2017} and the memory effect \cite{Judkewitz2015,Osnabrugge2017}. It has been used to generate nonlinear signals \cite{Mounaix2018} and for point-scan nonlinear imaging \cite{DeAguiar2016}. To obtain the TM of a scattering medium and thus, the linear relation between input and output fields, it is required to measure coherent fields in phase and amplitude by interferometric schemes with internal \cite{Popoff2010,Popoff2011,Tao2015,Yoon2015,DeAguiar2016,DeAguiar2017,Mounaix2017} or external reference beams \cite{Boniface2016,Mounaix2018,Kadobianskyi2018}. External references deliver flat wavefronts, but suffer from instability and make the setup more complex and less compact. The use of an internal reference circumvents this and furthermore benefits from an inherent coherence gating that preserves short pulses \cite{Mounaix2017}. However, there are two drawbacks when using an internal reference encoded onto a spatial light modulator (SLM). Firstly, some parts of its active area have to be sacrificed to the reference beam. Two sequential TM measurements using half of the SLM as reference and the other half for the modulated part and vice versa allow to acquire a TM for the full active area of the SLM, however in a more time consuming scheme \cite{Lee2016}. Secondly, and this is the problem discussed in this letter, the internal reference part also propagates through the scattering medium and manifests itself as a static speckle. In the dark regions of this speckled reference where the field amplitude is very low or vanishing, the interferometric measurements are biased by noise or cannot be performed at all. This leads to an incomplete measurement of the TM which is particularly disturbing for nonlinear imaging (e.g. two photon fluorescence or second harmonic generation (SHG)) because the linear incident field is squared or at higher orders, which leads to enhanced inhomogeneities when the focus is scanned. A scaling factor has to be introduced to partly correct for it, however with a high sensitivity to noise \cite{DeAguiar2016}. To correct for this effect, it had been shown that shifting the reference speckle can be implemented using a binary amplitude modulation with a digital mirror device \cite{Tao2015}; however, this shift has to be done for each individual pixel of the output field, which increases the measurement time. Likewise, a TM measurement without reference based on phase retrieval algorithms had been demonstrated with the drawback of lengthy computations \cite{Dremeau2015}.

\begin{figure}[tbp]
\centering
\includegraphics[width=\linewidth]{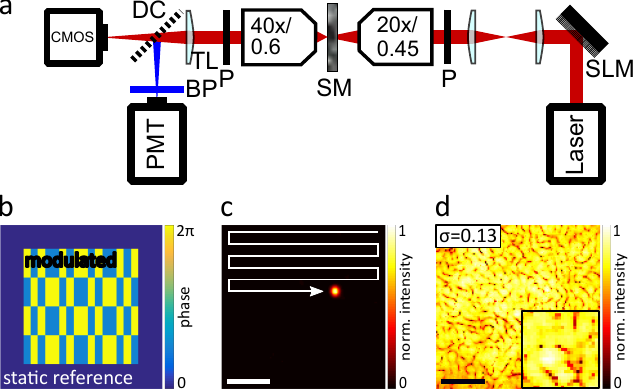}
\caption{a) Wavefront shaping setup with nonlinear signal detection. SLM: spatial light modulator; P: polarizer; SM: scattering medium; TL: tube lens; DC: dichroic mirror (LP758); BP: bandpass filter (414/46); PMT: photomultiplier tube. b) The SLM is divided into a center part with Hadamard bases projected onto it, and a static reference part. c) Principle of line scanning of the refocus with the  TM. d) A scan leads to a speckled intensity distribution as the TM is measured incompletely due to the speckled reference field (inset is a 4x zoom of the center). $\sigma$ is the standard deviation of the intensity (normalized to its maximum value) of the entire image. Scale bars: 10 µm.}
\label{fig:setup}
\end{figure}

In this letter, we present two direct solutions to correct for the TM measurement heterogeneities and provide a corrected TM that can be exploited for large FOV imaging. The first one takes advantage, in thin scattering media, of angular memory effect correlations. We demonstrate the possibility to extract these correlations from a single TM measurement, and exploit them to correct it \textit{a posteriori} with minimal computational effort. As memory effect correlations are low in thick biological tissues or highly scattering media, we propose a second alternative where the TM is measured with different phase masks that generate complementary reference speckle fields \cite{Gateau2017}, allowing a full output field TM composition. With this corrected TM at hand, we are able to generate nonlinear signals in FOVs above the memory effect range (e.g. a few tens of $\mu$ms) through fixed mouse spinal cord tissue slices, which exhibit high scattering \cite{Filatova2017}.
The use of a co-propagating reference beam keeps the nonlinear microscopy setup relatively simple as schematically depicted in Fig.\ref{fig:setup}a. The incident beam comes from a 140 fs pulse laser at 80 MHz repetition rate (Chameleon Ultra II, Coherent) and 790 nm wavelength. The linear signal used to measure the TM is recorded on a CMOS camera (BFLY-U3-23S6M-C, FLIR) whereas the nonlinear generated signal is recorded by a photomultiplier tube (R9110, Hamamatsu). The SLM (HSP256-0785, BNS) plane is conjugated to the back aperture of the excitation objective such that each pixel on the SLM corresponds to one input $k$-vector on the scattering medium. The elements of the TM $t_{mn}$ link the input fields from the SLM \(E_{in}\) to the output fields \(E_{out}\) behind the scattering medium, \(E_{out,n}=\sum_m^M t_{mn} E_{in,m}\) with $m$ the SLM pixel index, $n$ the camera pixel index and $M$ the total number of pixels on the SLM used to correct the wavefront \cite{Popoff2011}. To access the TM elements, the output field is measured using a step interferometric scheme which allows to access both its amplitude and phase.The outer part of the SLM serves as an internal static reference while the center active area of the SLM \(E_{mod}\) is modulated using a set of 1024 Hadamard bases (Fig.\ref{fig:setup}b), which are subsequently brought into a canonical representation by a unitary transformation. For each Hadamard base, the intensity $I_{out,n}(\alpha)$ is recorded with $\alpha$ the modulation phase, stepped from 0 to 2$\pi$:
\begin{equation} \label{eq:modulation}
\begin{split}
I_{out,n}(\alpha)=|\boldsymbol{E}_{ref,n}+\sum_m^M \boldsymbol{t}_{mn} \boldsymbol{E}_{mod,m}(\alpha)|^2=|\boldsymbol{E}_{ref,n}|^2 \\
+|\sum_m^M \boldsymbol{t}_{mn} \boldsymbol{E}_{mod,m}(\alpha)|^2
+2 Re(\boldsymbol{E}_{ref,n} \sum_m^M \boldsymbol{t}_{mn} \boldsymbol{E}_{mod,m}(\alpha))
\end{split}
\end{equation}
with $\boldsymbol{E}_{ref,n}$ the output reference field created by the static part of the SLM. 
The harmonic modulation of Eq. \ref{eq:modulation} gives the complex output field which is composed by the product of the reference field with the modulated field. Because the static reference field also propagates through the scattering medium, it appears as a static speckle with characteristic bright speckle grains and dark spots. At the proximity of these dark spots, the accuracy of the interferometric measurement decreases until failing to measure the output field. Hence, the TM obtained is incomplete. To visualize the consequence of this effect, the measured TM is employed to scan a refocus behind the scattering medium, as sketched in Fig.\ref{fig:setup}c. We map the maximum refocus intensity for each output pixel in Fig.\ref{fig:setup}d. The inhomogeneity of this refocus intensity map is the result of the incomplete TM measurement that is induced by the reference speckle amplitude. In fact, Fig.\ref{fig:setup}d depicts the reference speckle amplitude itself multiplied by a scaling factor \cite{DeAguiar2016}.

To correct for the low amplitude regions of the reference speckle, we propose two different methodologies for different scattering regimes: Firstly, at depths were light is scattered only a few times, the memory effect of the medium is larger than or comparable to the speckle grain size. Thus, the translation of a refocus from a bright reference spot towards a dark spot can be done efficiently with phase ramps. Equivalently, there is a simple phase relation between the TM elements that relate the input fields required to focus in these two neighbor regions. The phase values of the missing TM terms where the reference amplitudes are low can thus be replaced with phase solutions from neighboring bright reference spots, plus an additional phase ramp that corresponds to the displacement to these neighbor regions. We developed an algorithm that performs this task automatically and within less than a second with a standard processor. This is done \textit{a posteriori} once the TM measurement is finished and does not require extra measurement steps.
\begin{figure}[tb]
\centering
\includegraphics[width=\linewidth]{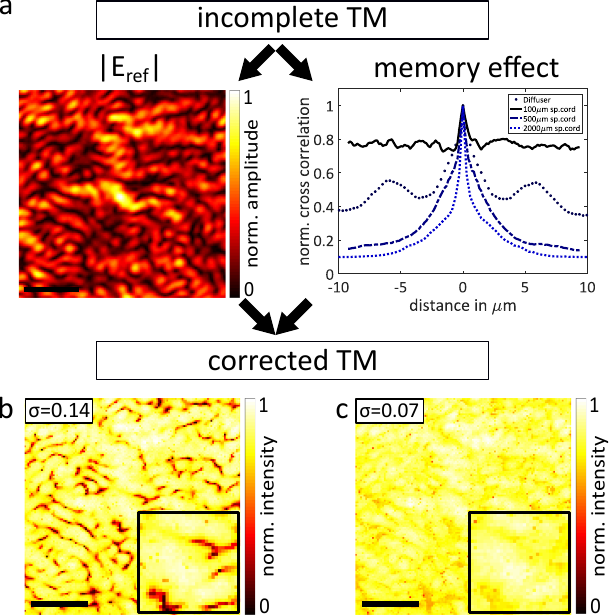}
\caption{a) Schematic principle of the TM correction based on the memory effect. The reference field amplitude and the memory effect are extracted from a TM measurement. The memory effect is represented for different samples (see text). b) Refocus intensity scans obtained with the uncorrected TM through a 500 µm thick mouse spinal cord tissue. c) Similar scan with the corrected TM. $\sigma$: standard deviation of the image intensity. Scale bar: 5 µm.}
\label{fig:SHTM}
\end{figure}
The algorithm exploits the intrinsic properties of the TM to extract two parameters, from a single TM measurement: first, the amplitude of the reference field and second, the memory effect of the medium (Fig.\ref{fig:SHTM}a). To extract the reference amplitude $|\boldsymbol{E}_{ref}|$ from the TM measurement, we take advantage of the fact that the reference field stays static during the projection of all orthogonal Hadamard bases. The interference term of Eq. \ref{eq:modulation} emphasizes that the measured TM elements $\boldsymbol{t}_{obs,mn}=\boldsymbol{t}_{mn}\boldsymbol{E}_{ref,n}$ are weighted by the reference speckle \cite{Popoff2011}. When summed in amplitude over all input pixels or bases, this term becomes:
\begin{equation} \label{eq:refampl}
\left\langle\boldsymbol|{t}_{obs,mn}|\right\rangle_{m} =\left\langle|\boldsymbol{t}_{mn}|\right\rangle_{m}
|\boldsymbol{E}_{ref,n}|
=const. \cdot |\boldsymbol{E}_{ref,n}| 
\end{equation}
where $const.$ is a spatially homogeneous constant background ($const.$) across the output field of view, since each input $m$ mode gives a speckle of same mean intensity \cite{Popoff2011}. The remaining spatially varying term is the reference amplitude. A normalization with its maximum value thus reveals the spatial amplitude profile of the reference speckle, as depicted in Fig.\ref{fig:SHTM}a. In our experiments 256 Hadamard bases suffice for a good estimation of the reference field amplitude.
The second step is to evaluate the extent of the memory effect in the medium. To do so, previous works have elaborated speckle correlation methods that can be time consuming \cite{Schott2015,Judkewitz2015}. In the present study, we propose an \textit{in situ} estimation purely based on one TM measurement, which contains its intrinsic spatial correlation properties. These correlations can be read-out by comparing input phase masks that refocus light in distinct output pixels. The subtraction of two complex wavefronts for such distinct output pixels (within the memory effect range) would reveal a linear phase ramp if a phase unwrapping algorithm was run. As we only want to measure correlations between the phase maps, we calculate the cross correlation of their Fourier transforms. It is calculated in horizontal and vertical directions and averaged over both dimensions. The results are shown in Fig.\ref{fig:SHTM}a for a diffuser, and for fixed mouse spinal cord tissue slices of thicknesses 100 µm, 500 µm and 2000 µm. Those results are comparable to a separate measurement where speckle decorrelations are measured upon applying input phase ramps \cite{Schott2015}. In particular, the memory effect in spinal cord tissues is considerably reduced at thicknesses above 500 µm, with a memory range below 5 µm in the image plane. In contrast, a 100 µm thick tissue (which is of the order of a scattering mean free path) affects poorly the beam propagation properties, such that the memory effect stays very large.
With the reference speckle map at hand we are able to determine the parts of the TM that need to be corrected, and the phase ramp extent required to perform such correction.
The algorithm used for the TM correction works as follows: we set a threshold of the normalized reference amplitude (with respect to its maximum value in the whole image, typically 0.3) below which the TM elements need to be corrected. Then, each corresponding row of the TM (that represents an input phase mask for refocusing) is replaced by one in its vicinity. We define a search radius around this element for which the reference amplitude is highest, take the phase mask that is required to refocus light in that particular output pixel, and add a linear phase ramps that shifts the focus towards the initial canceled output pixel. Depending on the memory effect range, the search radius is defined, typically not larger than one or two speckle grains. A conventional TM refocus scan is depicted in Fig.\ref{fig:SHTM}b for a 500 µm thick mouse spinal cord tissue, which shows visible heterogeneities in the retrieved image quantified by a high standard deviation of the image intensity. In comparison, the corrected TM shows a strong improvement of the image homogeneity (Fig.\ref{fig:SHTM}c). Even in the high scattering regime taking place in this sample, the TM correction replaces efficiently the low amplitude parts of the reference field speckle.
\begin{figure}[tb]
\centering
\includegraphics[width=\linewidth]{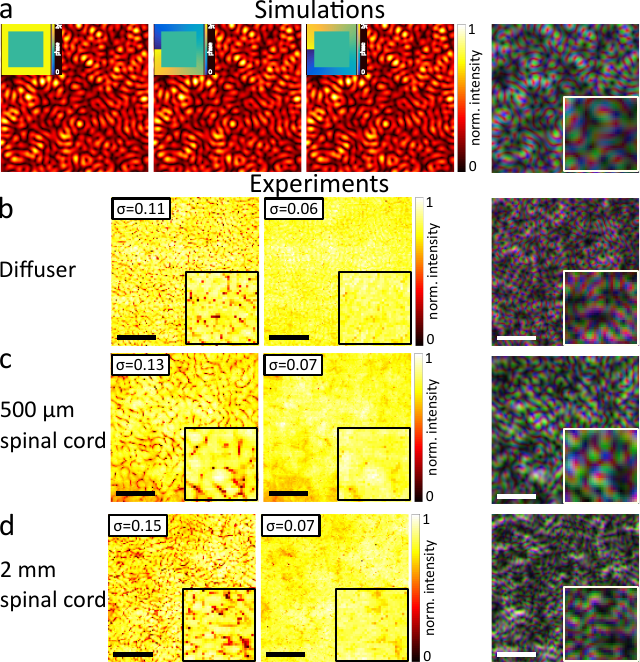}
\caption{a) Simulation of the projection of a flat phase (left), two spiral phase masks of +1 (middle) and -1 (right) topological charges on the reference part of the SLM (masks used for calculations and experiments are of circular shape). Right: complementary reference speckles encoded in RGB using the three phase masks shown in a). b-d) Refocus normalized intensity scans obtained with the uncorrected TM (left) and the complementary TM (middle) for different scattering media. Right: measured complementary reference speckles. Scale bars in b,d): 10 µm, in c): 5 µm.}
\label{fig:CplRefSpcklTM}
\end{figure}

With an increasing thickness of the medium, the memory effect is strongly reduced and the aforementioned method is less and less effective. For thick biological tissue or highly scattering media, we introduce another concept that is not relying on intrinsic properties of a TM measurement anymore, but rather on the alteration of the reference speckle in a controlled fashion. As described by Gateau et al. \cite{Gateau2017}, spiral phase patterns turn a speckle produced by a flat wavefront into complementary speckles with interchanged bright and dark spots. To exploit this property, three independent TM measurements are performed with three different phase patterns on the reference area: a flat wavefront and two spiral phase masks of opposite topological charges (+1 and -1) (insets of Fig.\ref{fig:CplRefSpcklTM}a). Each of those three TM measurements have complementary reference field amplitudes (as shown in a simulation in Fig.\ref{fig:CplRefSpcklTM}a where the SLM pixels where multiplied with a random phase pattern that mimicked a scattering medium and Fourier transformed to visualize the speckle pattern) of which we compose a new "complementary TM" by picking the TM values of the brightest reference speckle grains of either of those three TMs.
We tested this method on three types of scattering samples: a diffuser, a 500 µm and a 2 mm fixed mouse spinal cord tissue. The refocus intensity maps of the uncorrected TM and the complementary TM are compared in Fig.\ref{fig:CplRefSpcklTM}b,c and d. In all cases, the obtained reference speckles using the three different phase masks clearly exhibit complementary features (Fig.\ref{fig:CplRefSpcklTM}, right). We observe a homogeneous refocus scan over a large FOV ($\sim$40x40 µm) for all three scattering regimes, even in the case of 2 mm thick spinal cord tissue which has a very low memory effect (Fig.\ref{fig:SHTM}a). The 500 µm thick tissue marks a transition where the memory effect is a few µms (Fig.\ref{fig:SHTM}), with similar performances in the corrected TM (Fig.\ref{fig:SHTM}c) and complementary TM (Fig.\ref{fig:CplRefSpcklTM}c). Despite its more lengthy procedure (the TM needs to be measured three times), the complementary TM leads to higher homogeneity in the refocus scan independently on the scattering medium properties. Note that using a higher number of complementary references (e.g. higher order spiral phases), leads to an extra improvement of the scan homogeneity. This is however at the price of a longer measurement. At last, other schemes than spiral phases can be used, which lead to similar results, such as random phase masks or phase shifted segmented phase patterns.
\begin{figure}[tb]
\centering
\includegraphics[width=\linewidth]{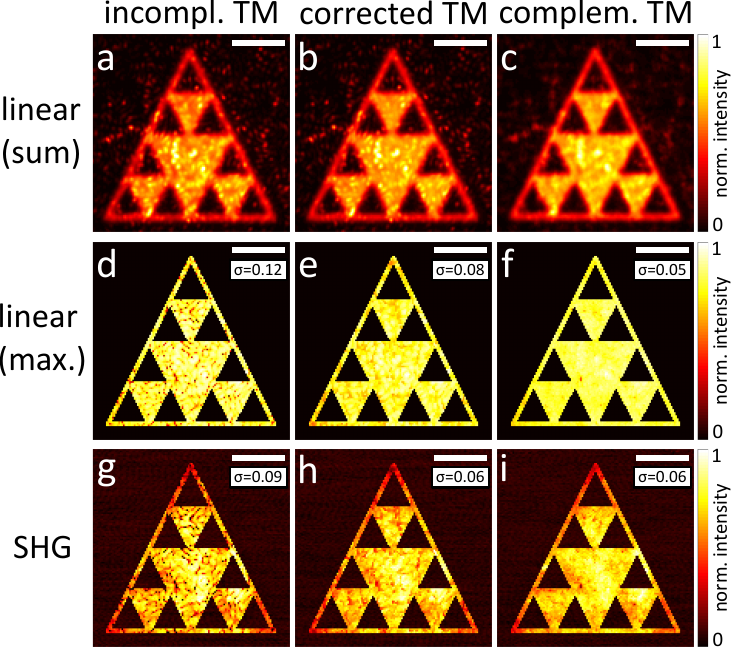}
\caption{Comparison of refocus homogeneity with three different TM in a $\sim$40x40 µm FOV for the sum of all linear refocii a)-c), the maximum intensity for each refocus d)-f) and the nonlinear signal (SHG) generation g)-i). The refocus is generated in a BBO crystal through 500 µm fixed spinal cord tissue light. Scale bars: 10 µm.}
\label{fig:NLImaging}
\end{figure}
We finally compare the impact of the uncorrected TM, the corrected TM (Fig.\ref{fig:SHTM}) and the complementary TM (Fig.\ref{fig:CplRefSpcklTM}) on the homogeneity of nonlinear second harmonic generation (SHG) across the output field through a 500 µm spinal cord tissue. A desired pattern of 30 µm full size made of triangular shapes (Fig.\ref{fig:NLImaging}) was used to focus the beam point-by-point in a SHG-active Beta Barium Borate (BBO) crystal of 500 µm thickness, at a distance of ~400 µm behind the spinal cord tissue.
Fig.\ref{fig:NLImaging}a-c show the sum of the full camera linear intensity during the refocus scan. The spatial refocus resolution is visible and reaches about 0.9 µm, which is limited by the collection objective numerical aperture. The background inhomogeneity is due to the static remaining reference speckle background, which sums up into a non vanishing pattern although still negligible compared to the refocus intensity. Fig.\ref{fig:NLImaging}d-f show the maximum refocus intensity per pixel. In both linear refocus representations, a gradual homogeneity improvement is visible from the uncorrected TM to the corrected TM and finally to the the complementary TM, which exhibits the lowest intensity standard deviation value. For the complementary TM, the refocus scan is not only more homogeneous but also of higher intensity, which illustrates the more optimal character of the method. The same trend is observed, as expected, in the SHG image depicted in Fig.\ref{fig:NLImaging}g-i. A small degree of remaining heterogeneity is observed for the complementary TM correction (Fig.\ref{fig:NLImaging}i), due to the fact that SHG signals are proportional to the square of the linear refocus intensity, which enhances intensity deviations. The SHG signals also drops at the border of the outer triangular structure, which we attribute to an increase in pulse broadening for off-axis refocusing \cite{Small2009,Small2012}. Note that very similar results were obtained in a 2 mm thick spinal cord tissue, using an incident power of about an order of magnitude higher to compensate for the lower refocus efficiency.

We have shown how to efficiently correct for heterogeneities that appear in a TM measurement due to the presence of scattered internal reference fields. For highly scattering samples, the complementary TM is visibly the most reliable approach to generate nonlinear signals homogeneously across large FOVs. This provides promising prospectives for the use of internal references in scattering media in general.

\textbf{Acknowledgments.} The authors thank S. Sivankutty for discussions and B. El Waly and F. Debarbieux (INT, Marseille, France) for the preparation of mouse spinal cord samples.

\textbf{Funding information.} This work has been supported by the contracts ANR-15-CE19-0018-01 (MyDeepCARS), and A*MIDEX NEUROPHOTONICS of the Aix Marseille University.

% Bibliography
\bibliography{CompletingTM}

% Full bibliography added automatically for Optics Letters submissions; the following line will simply be ignored if submitting to other journals.
% Note that this extra page will not count against page length
% \bibliographyfullrefs{CompletingTM}

\end{document}